\documentstyle [twocolumn,epsf]{mn}
\newcommand{\mincir}{\raise
-3.truept\hbox{\rlap{\hbox{$\sim$}}\raise4.truept\hbox{$<$}\ }}
\newcommand{\magcir}{\raise
-3.truept\hbox{\rlap{\hbox{$\sim$}}\raise4.truept\hbox{$>$}\ }}
\newcommand{\minmag}{\raise
-3.truept\hbox{\rlap{\hbox{$<$}}\raise5.truept\hbox{$<$}\ }}
\newcommand{\be}{\begin{equation}}
\newcommand{\ee}{\end{equation}}
\newcommand{\ba}{\begin{eqnarray}}
\newcommand{\ea}{\end{eqnarray}}
\newcommand{\brr}{\begin{array}}
\newcommand{\err}{\end{array}}
\newcommand{\bc}{\begin{center}}
\newcommand{\ec}{\end{center}}

\newcommand{\hm}{\,h^{-1}{\rm Mpc}}

\renewcommand{\baselinestretch}{1.0}

\title[Shape statistics]
{Shape statistics of Sloan Digital Sky Survey superclusters}
\author[Spyros Basilakos ]{Spyros Basilakos$^{1}$. \\
\vspace{0.1cm}
$^1$ Institute of Astronomy \& Astrophysics, National Observatory of Athens, 
I. Metaxa \& V. Pavlou, Palaia Penteli, 15236 Athens, Greece \\
}

\begin{document}

\maketitle

\begin{abstract}
We study the supercluster shape 
properties of the recently compiled
SDSS cluster catalog using an approach
based on differential geometry. We detect superclusters by
applying the percolation algorithm to observed cluster
populations, extended out to $z_{\rm max}\leq 0.23$ in order
to avoid selection biases.
We utilize a set of shapefinders in order to study the
morphological features of superclusters with $\geq 8$ cluster members
and find that filamentary morphology is the dominant supercluster 
shape feature, in agreement with previous studies.

{\bf Keywords:} cosmology: theory - clusters - superclusters: general -
large-scale structure of universe -  Optical: clusters
\end{abstract}

\vspace{1.0cm}

\section{Introduction}
The launch of the recent observational data 
has brought great progress in understanding the
cosmic structure formation pattern. From the large scale structure point
of view it has been shown that cluster of galaxies are not 
randomly distributed but tend to aggregate 
in larger groups, the so called superclusters (cf. Bahcall 1988).
Since they have been seeded by density
perturbations of the largest scale ($\sim 100\hm$), they therefore constitute
objects with which one can study the details of the fluctuations that
gave rise to cosmic structures 
(cf. West 1989; Einasto et al. 1997). In this framework, 
we can extract useful information regarding the 
evolution of the large-scale 
structure of the universe 
and test cosmological models within the framework of hierarchical structure 
formation scenario (Bahcall \& Soneira 1984; 
Bahcall 1988; Frisch et al. 1995).

Indeed many authors have confirmed that the large scale 
clustering pattern of galaxies is described well by a filamentary distribution
(Zel'dovich, Einasto \& Shandarin 1982;
de Lapparent, Geller \& Huchra 1991; 
Einasto et al. 2001). However, only recently 
(Sathyaprakash et al. 1998a; Basilakos, 
Plionis \& Rowan-Robinson 2001; Kolokotronis, Basilakos \& Plionis 2002) 
was any significance given to cosmological inferences from 
supercluster shape statistics claiming that a 
low matter density ($\Omega_{\rm m}=1-\Omega_{\Lambda}=0.3$) 
flat cosmological model 
($\Lambda$CDM) fits the observational results at high 
significance level. Sathyaprakash et
al. (1998a) and Basilakos et al. (2001) have used infrared 
galaxy samples (1.2 Jy and PSCz,
respectively), whereas Kolokotronis, et al. (2002) 
considered Abell/ACO clusters. 

A variety of geometrical and topological methods have been
developed and applied to observational data 
in order to give qualitative and quantitative description of large scale
structure (cf. Weinberg, Gott \& 
Melott 1987; Coles \& Plionis 1991; Mecke et al. 1994;
Sahni \& Coles 1995; Yess \& Shandarin 1996; 
Kerscher et al. 1997; Kerscher et al. 2001a, b; Hoyle, Vogeley \& Gott 2002;
Hoyle et al. 2003; Jatush et al. 2003 and references therein).
Recently, a new technique was obtained by Sahni, Sathyaprakash \& 
Shandarin (1998) based on the differential 
geometry (Minkowski functionals), in order to 
describe in detail the geometrically and topologically complex
features of large scale structure. 
In its general form the Minkowski functionals have been shown to 
provide a better approximation to complex structures than the simple
semi-axes procedures (Sathyaprakash, Sahni \& Shandarin 1998b).
For example, in cases where the matter distribution 
is one-dimensional but not along straight line, one is bound to get somewhat
smaller signal for filamentarity from the Babul \& Starkman (1992) and 
Luo \& Vishniac (1995) statistics (semi-axes method). This problem 
is even more true for two dimensional structures. 

The first time this new shapefinder technique was applied
to astronomical data was in Basilakos et al. (2001), where it 
ascertained that the prominent
feature of the large scale structures we see today is filamentarity, as it
has also been observed in N-body simulations of gravitational clustering
(Sathyaprakash et al. 1998b and references therein).
In this paper we utilize the recently completed SDSS CE 
cluster catalog (Goto et al. 2002) in order:
(i) to study whether we can reliable identify SDSS superclusters and 
measure their shape and size distribution and (ii) 
whether or not superclusters verify the dominance of filamentarity as 
the basic trait of large scale structure.
A different analysis applied by Einasto et al. (2002), who identified 
clusters and 43 superclusters (high density regions) from the smoothed
galaxy density field of the SDSS Early Data Release (Stoughton 
et al. 2002) catalog. The plan of the paper is as follows. The 
observed dataset is presented in section 2. 
In section 3, we give a brief report on the 
method used to investigate supercluster shape 
properties and comment on the systematic effects introduced in our
analysis. In section 4, we present the
morphological parameters of the SDSS
superclusters data and finally in section 5 we draw our conclusions.

\section{The SDSS cluster catalogue}
In the present analysis we use the recent SDSS CE cluster catalog 
(Goto et al. 2002), which contains 2770 and 1868 galaxy clusters 
in the North ($145.1^{\circ} <RA< 236.0^{\circ}$, 
$-1.25^{\circ} <DEC< 1.25^{\circ}$) and South 
($350.5^{\circ} <RA< 56.61^{\circ}$, 
$-1.25^{\circ} <DEC< 1.25^{\circ}$) 
slices respectively, 
covering an area of $\sim 400$deg$^2$ in the sky.
Redshifts are converted to proper distances 
using a spatially flat cosmology with 
$H_{\circ}=100h\,$km$\,$s$^{-1}\,$Mpc$^{-1}$ and 
$\Omega_{\rm m}=1-\Omega_{\Lambda}=0.3$.
\begin{figure}
\mbox{\epsfxsize=8cm \epsffile{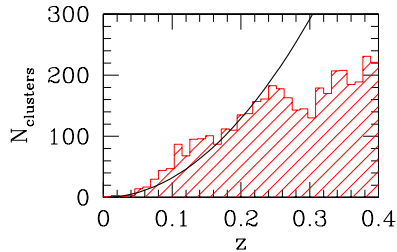}}
\caption{The estimated (histogram) and the expected (line) number 
of the SDSS clusters as a function of redshift.}
\end{figure}
In figure 1, we present the 
estimated (histogram) and the expected (line) number 
of the SDSS clusters as a function of redshift. 
It is obvious that up to redshift $z_{\rm max}\le 0.23$ we have 
a volume limited sample due to the fact that the number of the 
SDSS clusters is proportional to $r^{3}$.
We have compare the two distributions (up to 
$z_{\rm max}\le 0.23$) via a standard Kolmogorov-Smirnov (KS) 
statistical test and the corresponding probability of 
consistency between model and observations is ${\cal P}_{\rm KS}\simeq0.43$.
Therefore, the redshift cutoff corresponds to a limiting 
distance $r_{\rm max}\le 653h^{-1}\,$Mpc.
This subsample contains 1690 entries in the North and South
slices respectively. 

In order to investigate further the latter result
we need to compute the number density of the cluster sample
as a function of redshift. We do so by using five 
equal volume shells ($\delta V\approx 2.4\times 10^{6}\,h^{-3}\,$
Mpc$^{3}$) up to $z_{\rm max}\le 0.23$ and calculate,
as a function of redshift, the cluster space density 
(continuous line in figure 2). 
While, the corresponding average-over-shells density 
for the SDSS subsample is 
$n_{\rm SDSS} (\le z_{\rm max})\simeq 1.26(\pm 0.32) 
\times 10^{-4}h^{3}\,$Mpc$^{-3}$ (see dashed line in figure 2)
giving rise to intercluster separation of order
$\sim 20 \pm 1.47$ $h^{-1}$Mpc.
. 
\begin{figure}
\mbox{\epsfxsize=8cm \epsffile{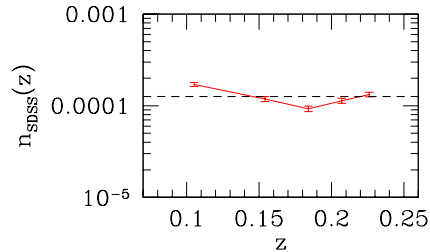}}
\caption{The SDSS cluster space density (continuous line) estimated
in equal-volume shells with its 
Poissonian uncertainty. While the dashed line is the average
$n_{\rm SDSS}(z\le 0.23)\simeq 1.26 
\times 10^{-4}h^{3}\,$Mpc$^{-3}$.}
\end{figure}

\section{Shape Statistic}

\subsection{SDSS Superclusters}

Identifying real superclusters is a difficult problem in
general, due to observational selection function, 
that can significantly affect the visual structure of
superclusters and distort the true pattern.
Our supercluster finding algorithm consists of the main steps: 
The supercluster catalogs constructed by utilizing, in supergalactic 
coordinates, a constant size neighborhood 
radius, i.e. the percolation radius
(Zel'dovich et al. 1982).
In particular, we place a sphere of a certain size around each
cluster and then find all neighboring spheres having an overlap region.
Then, all mutually linked clusters are joined together to form groups (with
$k$ members), and the
groups with more than $k\ge 8$ clusters are 
identified as candidate rich superclusters.
Of course, we choose the optimal percolation radius, 
by repeating the procedure by 
successively increasing the size of the sphere. At the end, we identify the 
radius that yields the maximum number of superclusters, which obviously 
occurs before the percolation of the superclusters themselves. Performing many 
tests we select a percolation radius $R_{\rm pr}\simeq 26h^{-1}$Mpc
following the same criteria 
to those of Einasto et al. (2001; Kolokotronis et al. 2002
and references therein). To this end this procedure 
produced a list of 57 rich superclusters.

\subsection{Test for systematic errors}
In this section we investigate the number  
of the random clumps revealed from the percolation procedure.
In particular, we run a large number of Monte-Carlo simulations in 
which we destroy the intrinsic SDSS clustering 
by randomizing the supergalactic coordinates of the clusters while 
keeping their distances and therefore their selection function unchanged. 
On this intrinsically random cluster distribution, 
we apply the procedure described before and identify the expected
random superclusters, $N_{\rm ran}$, which are due to our 
supercluster-identification method itself.

In figure 3 (top panel) we plot for the different percolation radii, 
the probability of detecting real clusters in the SDSS data, defined as
${\cal P} = 1-N_{\rm ran}/N_{\rm SDSS}$  
as well as the number of real, $N_{\rm SDSS}$ (continuous line), 
and random, $N_{\rm ran}$ (dashed line), superclusters. 
In the presence of negligible systematic biases in our
method and data, the above selection process should result in ${\cal P}
\simeq 1$. 
In other words, the closer ${\cal P}$ is to 1 the less likely to reveal
random superclusters.
\begin{figure}
\mbox{\epsfxsize=8cm \epsffile{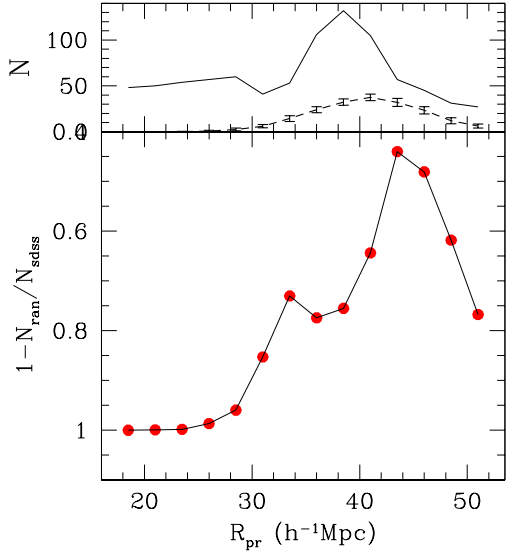}}
\caption{Statistical significance of our supercluster detection 
procedure as a function of percolation radius $R_{\rm pr}$. 
Up panel shows the number of real, $N_{\rm SDSS}$ (continuous line), 
and random, $N_{\rm ran}$ (dashed line), superclusters respectively.
Top panel shows the corresponding probability of detecting 
real superclusters in the SDSS data. Note that this plot 
accounts for clumps with at least eight cluster members.}  
\end{figure}
To this end from figure 3 it is evident that with a threshold of
$26h^{-1}$Mpc for the percolation radius, 
we find an average of 0.4 random superclusters in the mock SDSS 
realizations.
Thus the superclusters we detect in the SDSS with $R_{\rm pr}=26h^{-1}$Mpc
are significant at the $\ge 99\%$ confidence level. This is why we 
set this percolation threshold.

\subsection{The Supercluster Global Geometry}
Shapes are estimated for those superclusters that 
consist of 8 or more clusters 
\footnote{In Kolokotronis et al. (2002), we have tested the 
performance of the shape method using 
a large set of Monte Carlo simulations (see their section 3.3) 
and they found that structures with more that 8 
members (points) are well described by the our shape method.}, utilizing 
the moments of
inertia ($I_{ij}$) method to fit the best triaxial
ellipsoid to the data (cf. Plionis, Barrow \& Frenk 1991). We diagonalize 
the inertia tensor 
\begin{equation}\label{eq:diag}
{\rm det}(I_{ij}-\lambda^{2}M_{3})=0 \;\;\;\;\; {\rm (M_{3} \;is \; 
3 \times 3 \; unit \; matrix) } \;,
\end{equation}
obtaining the eigenvalues $\alpha_{1}$, $\alpha_{2}$, 
$\alpha_{3}$ (where $\alpha_{1}$ is the semi-major axis) 
from which we define the shape of the configuration since, the
eigenvalues are directly related to the three principal axes 
of the fitted ellipsoid 
\be
{\mbox{\boldmath$r$}} (\theta,\phi)=a_{1}\,{\rm sin}\theta\,{\rm cos}\phi\,
{\mbox{\boldmath${\hat i}$}}+
a_{2}\,{\rm sin}\theta\,{\rm sin}\phi\,{\mbox{\boldmath${\hat j}$}}+
a_{3}\,{\rm cos}\theta\,{\mbox{\boldmath${\hat k}$}} \;\;, 
\label{eq:parform}
\ee
having volume 
$V=\frac{4\pi}{3} \alpha_{1} \alpha_{2} \alpha_{3}$
and $0\le \phi \le 2\pi$, $0\le \theta \le \pi$.   
The shape statistic procedure, that we use, is based on a differential
geometry approach, introduced by Sahni et al.(1998). In this paper 
we review only some basic concepts.

Furthermore, having identified superclusters with $k\ge 8$ members 
a set of three shapefinders are defined having dimensions of length;
${\cal H}_{1}=V S^{-1}$, ${\cal H}_{2}=S C^{-1}$ and ${\cal H}_{3}=C$, 
with $S$ the surface area and $C$ the integrated mean curvature.
In this framework, based on these, the dimensionless shapefinders 
$K_{1}$ and $K_{2}$ can be defined as follows:

\be
K_{1}=\frac{ {\cal H}_{2}-{\cal H}_{1} }{ {\cal H}_{2}+{\cal H}_{1} } 
\ee

and

\be
K_{2}=\frac{ {\cal H}_{3}-{\cal H}_{2} } { {\cal H}_{3}+{\cal H}_{2} } \;\;, 
\ee
where $K_{1,2}\le 1$ by definition and 
normalized to give ${\cal H}_{i}=R$ ($K_{1,2}=0$) for a sphere of radius $R$.
The above shape vector 
${\bf K}=(K_{1},K_{2})$ 
characterizes the shapes of topologically 
non-trivial cosmic objects according to the following classification:
(i) pancake-like ellipsoids for $K_{1}>K_{2}$ or ${\cal R}=K_{1}/K_{2}>1$;
(ii) filament-like ellipsoids for $K_{1}<K_{2}$ or ${\cal R}=K_{1}/K_{2}<1$; (iii) 
triaxial for $K_{1}=K_{2}$ or ${\cal R}=K_{1}/K_{2}=1$ and (iv)
spheres for $\alpha_{1} =\alpha_{2} =\alpha_{3}$ and 
thus $(K_{1},K_{2})=(0,0)$. 
Note that for ideal filaments ${\bf K}=(0,1)$, pancakes ${\bf K}=(1,0)$, 
triaxial structures ${\bf K}=(1,1)$ 
and spheres ${\bf K}=(0,0)$.

\begin{figure}
\mbox{\epsfxsize=8.5cm \epsffile{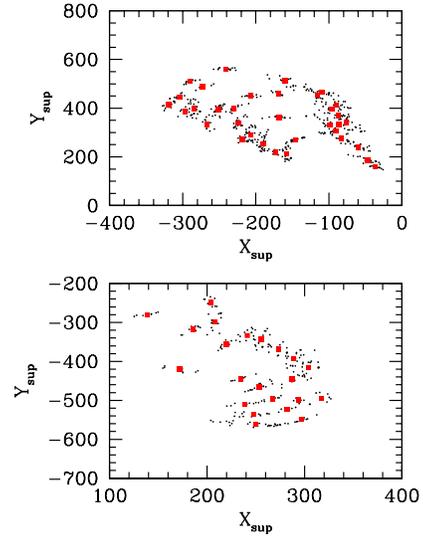}}
\caption{Two dimensional ($Z_{\rm sup}=0$) whole sky 
map of the 57 SDSS superclusters
(squares). Dots denote the clusters associated with superclusters 
with $k\ge 8$. In this distribution we have detected 34 
superclusters in the North and 
23 in the South.} 
\end{figure}
For the quasi-spherical objects, $K_{1}$ and $K_{2}$ 
are both very small (order of $\sim 10^{-3}-10^{-4}$), 
and thus the ratio ${\cal R}=K_{1}/K_{2}$ 
measures the deviation from pure sphericity
(for further details see Basilakos et al. 2001).
It is interesting to mention that in the present work we didn't
find such small values for the $(K_{1},K_{2})$ [for further details 
see next section].

\section{Morphological Properties}
We investigate supercluster characteristics according to the observational
requirements and the definitions of shape statistic set in section 3.
Taking the optimum value of $R_{\rm pr}=26h^{-1}\,$Mpc for the
combined SDSS sample, we find 57 superclusters with $k\geq 8$.
Table 1, lists all the relevant information.
In figure 4, we show a 2D ($Z_{\rm sup}=0$) schematic representation 
of superclusters (squares) 
and the related cluster distribution (dots) for the 
SDSS sample. Note, that only superclusters
with $k\geq 8$ and their member clusters are plotted.

\begin{table*}
\caption[]{List of the SDSS superclusters using
$R_{\rm pr}=26h^{-1}\,$Mpc. The correspondence of the columns is as follows:
index number (the first 34 superclusters belong to the North and 
the rest of them in the South slice respectively), multiplicity, axes of 
triaxial ellipsoid, redshift, 
right ascension $RA$ and declination $DEC$ of the supercluster center, $K_{1}$ and $K_{2}$ are the
shapefinders, 
their ratio ${\cal R}=K_{1}/K_{2}$ and 
finally the morphological classification 
type. F 
denotes filaments, P is for pancakes and cT is close to triaxial. Note 
that the index B, wherever exists, means that
the specific entry is close to the boundaries 
of the SDSS cluster catalog. 
Finally, The $a_{1}$, $a_{2}$ and $a_{3}$ have units of $h^{-1}\,$Mpc.}

\tabcolsep 9pt
\begin{tabular}{cccccccccccc} 
\hline
Index & $k$ & $\alpha_{1}$ & $\alpha_{2}$ & $\alpha_{3}$ & $z$ & RA & DEC &$K_{1}$&$K_{2}$&${\cal R}$
& Type \\ \hline \hline 
   1  &  18 &  56.20&  24.15&  21.02&        0.083&   154.00&    0.20& 0.045&0.097&  0.46&F-B\\
   2  &  17&   34.79&  15.87&  12.69&        0.219&   151.45&   -0.07& 0.048&0.093&  0.52&F-B\\
   3 &   16&   42.80&  20.95&  15.71&        0.196&   149.62&    0.01& 0.050&0.084&  0.59&F-B\\
   4  &  15&   44.06&  19.04&  16.89&        0.106&   154.48&    0.47& 0.043&0.094&  0.46&F-B\\
   5 &   10&   36.23&  16.89&  12.59&        0.162&   149.78&   -0.04& 0.054&0.095&  0.56&F-B\\
   6 &    9 &  33.46&  15.13&  13.12&        0.072&   151.58&    0.15& 0.042& 0.086&  0.48&F-B\\
   7 &   14&   48.41&  22.58&  18.31&        0.174&   150.99&   -0.24& 0.046&0.087&  0.52&F-B\\
   8 &   21&   81.49&  35.79&  30.22&        0.207&   153.96&    0.07& 0.046&0.096&  0.48&F-B\\
   9 &   10&   46.67&  21.31&  18.05&        0.185&   151.47&   -0.36& 0.043&0.087&  0.49&F-B\\
  10 &   13&   29.85&  13.34&   8.80&        0.128&   160.11&    0.26& 0.072&0.117&  0.62&F\\
  11&    12&   64.82&  25.43&  19.10&        0.219&   161.52&    0.01& 0.064&0.136&  0.47&F\\
  12 &   10&   37.47&  16.39&  10.96&        0.140&   160.07&    0.11& 0.072&0.121&  0.60&F\\
  13 &   23&   91.56&  39.49&  35.15&        0.150&   154.03&   -0.14& 0.043& 0.094&  0.46&F-B\\
  14 &   24 &  79.67&  31.36&  28.34 &       0.116&   160.16 &   0.02& 0.048&0.113&  0.42&F\\
  15 &    8  & 37.02&  10.08&   9.07  &      0.185&   169.56  & -0.24& 0.068& 0.218&  0.31&F\\
  16 &   12  & 69.15 &  8.23&   2.12   &     0.139&   184.68   &-0.45& 0.375&0.610&  0.62&F\\
  17 &   11&   48.46 & 11.00&   2.36    &    0.219&   180.92 &   0.18& 0.430& 0.410&  1.05&cT\\
  18 &   10 &  43.75 &  8.08&   1.33     &   0.174&   184.71  & -0.23& 0.514& 0.481&  1.07&cT\\
  19  &  22  & 49.41 & 12.22&  10.13    &    0.105 &  199.04   &-0.08& 0.078& 0.261&  0.30&F\\
  20  &  12  & 42.01 & 14.24&   3.95   &     0.208  & 202.48  & -0.40& 0.315& 0.261&  1.20&P\\
  21   & 10  & 46.92 & 14.78&   6.72   &     0.196 &  200.96  & -0.41 &0.171& 0.251&  0.68&F\\
  22&    16  & 69.95 & 15.87&  11.76   &     0.162  & 204.20   &-0.16  & 0.093& 0.306&0.30&F\\
  23 &   22  & 75.40 & 15.70&  10.55    &    0.174&   214.47&    0.22 &  0.110& 0.350&0.31&F\\
  24  &  33  &109.75 & 11.87&   2.42     &   0.128 &  225.89 &  -0.14  & 0.454& 0.640& 0.71&F\\
  25 &   21  & 78.94 & 16.00&  11.65      &  0.151  & 215.41  & -0.25   &0.099& 0.350& 0.28&F\\
  26  &  30 &  84.15 &  9.36&   1.28       & 0.139&   222.32 &  -0.11 &  0.582& 0.640& 0.91&F\\
  27  &  26 &  65.28 &  8.71&   5.55 &       0.105 &  226.05 &  -0.12 &  0.130& 0.510& 0.25&F-B\\
  28  &  12 &  36.51 & 11.12&   2.00  &      0.208  & 219.10 &  -0.20  & 0.470& 0.311&1.50&P\\
  29  &  23 &  78.40 & 12.78&   6.92   &     0.196 &  224.02  &  0.25   &0.160& 0.460&0.34&F\\
  30  &  16 &  37.64 & 12.57&   6.75   &     0.219 &  227.71   &-0.25 &  0.130& 0.220&0.59&F-B\\
  31  &   9 &  22.89 &  8.52&   3.23    &    0.151 &  232.34   & 0.27  & 0.211& 0.220&0.96&cT-B\\
  32  &   9 &  33.08 & 11.50&   2.06     &   0.208 &  229.53   & 0.29  & 0.462&0.270&1.71&P-B\\
  33  &  12 &  44.35 & 10.41 &  8.02      &  0.185 &  231.34   & 0.13 &  0.088&0.290&0.30&F-B\\
  34  &  12 &  30.66 &  9.74  & 6.12       & 0.117  & 230.55   & 0.06  & 0.101&0.220&0.47&F-B\\
  35  &  29 & 121.03 & 17.30  &12.13&        0.219   &  2.86   & 0.11&  0.114& 0.474& 0.24&F\\
  36  &  10 &  58.86 & 13.48  & 9.95 &       0.163 &  357.00   &-0.02 &  0.093& 0.303&0.31&F-B\\
  37  &  16 &  79.83 & 14.40&   4.07  &      0.209  &   6.04   & 0.02  & 0.336&0.471&0.71&F\\
  38  &  11 &  44.19 & 10.74 &  3.01   &     0.197  &   7.00   & 0.06 &  0.330& 0.370&0.89&F\\
  39  &   8 &  37.58 & 12.97  & 9.31    &    0.175  &  17.16   & 0.03  & 0.077&0.180&0.44&F\\
  40  &  18 &  61.42 & 17.56  &10.72     &   0.186  &  20.94  & -0.25   &0.111&0.253&0.44&F\\
  41  &  10&   66.06 &  8.82  & 4.79      &  0.072  &  11.27  &  0.02 &  0.159& 0.530&0.30&F\\
  42  &  10 &  32.35 & 14.59  & 2.80       & 0.198  &  18.64  &  0.12 &  0.420&0.193&2.17&P\\
  43  &   8 &  46.06 &  8.43  & 7.45&        0.107  &  12.01  & -0.02 &  0.084&0.354&0.24&F\\
  44  &  11 &  36.85 & 12.42  & 4.51 &       0.220  &  18.29  &  0.28 &  0.240&0.250&0.96&cT\\
  45  &   8 &  35.27 &  7.06  & 1.20  &      0.209  &  18.72   & 0.19 &  0.502&0.453&1.11&P\\
  46  &  11 &  45.26 & 12.57 &  8.28   &     0.129   & 25.72  &  0.04 &  0.100&0.253&0.40&F\\
  47  &   9 &  39.60 & 14.05 &  2.30    &    0.220    &34.21  &  0.03 &  0.491& 0.266&1.84&P\\
  48  &  14 &  44.69 & 15.31 & 11.82     &   0.152 &   32.15   &-0.28 &  0.069&0.170&0.41&F\\
  49  &  15 &  71.83 & 14.17 &  7.63      &  0.140  &  41.25   & 0.09 &  0.151&0.393&0.38&F\\
  50  &   8 &  45.36 & 15.09 & 11.86       & 0.197   & 34.15   & 0.20 &  0.069&0.180&0.39&F\\
  51  &   9 &  37.99 &  6.19 &  2.93&        0.209 &   27.32   &-0.15 &  0.190&0.472&0.39&F\\
  52  &  19 &  36.46 & 10.30 &  1.57 &       0.197  &  44.74   & 0.14 &  0.527&0.340&1.55&P\\
  53  &  24 &  68.23 & 16.56 &  0.70  &      0.209   & 44.43   & 0.08 &  0.835& 0.402&2.08&P\\
  54  &  11 &  31.04 & 11.20 &  3.78   &     0.186 &   45.22   & 0.22 &  0.250&0.235&1.07&cT\\
  55  &  20 &  55.01 & 13.62 &  6.21    &    0.174  &  46.07   & 0.01 &  0.182& 0.330&0.55&F-B\\
  56  &   8 &  25.87 &  8.36 &  1.09     &   0.163   & 43.59   & 0.13 &  0.570& 0.300&1.90&P\\
  57  &   8 &  20.44 &  7.39 &  4.80      &  0.141   & 53.12   & 0.00 &  0.090& 0.173&0.51&F-B\\
\hline
\end{tabular}
\end{table*}

\begin{figure}
\mbox{\epsfxsize=8.0cm \epsffile{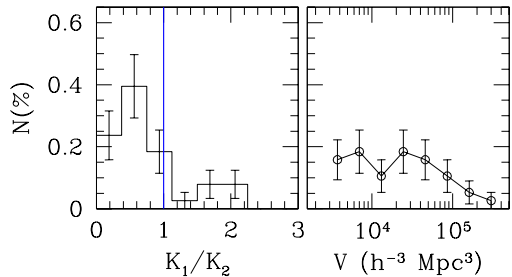}}
\caption{The shape and volume spectrum of the 
SDSS superclusters. The error-bars are estimated utilizing Poisson 
statistics} 
\end{figure}
As is evident from Table 1, there are 19 entries
which are near the boundaries of the SDSS cluster catalog. Therefore,
in order to avoid ill shape definitions, for the shape statistics, we
feel more secure to use superclusters which are away from the
boundaries. This selection procedure decreases the total number of
analysed entries to 38. 

Again from Table 1, it is obvious that there are no spherical superclusters in 
concordance with numerical N-body simulations of gravitational clustering and 
similar studies on observed data (Plionis, Valdarnini \& Jing 1992; 
Sathyaprakash et al. 
1998a, b; Sahni et al. 1998; Basilakos et al. 2001; 
Kolokotronis et al. 2002). 
We plot in figure 5 the shape-spectrum 
as well as the multiplicity function (volume spectrum). The shape 
spectrum shows 
a measure of the global SDSS superclusters geometry. It is obvious
that filamentary structures dominate our supercluster sample,
in agreement with previous studies 
(Sathyaprakash et al. 1998a,b; Basilakos et al. 2001; 
Kolokotronis et al. 2002). 
It is interesting to mention that 
the vast majority of the systems (71$\%$ or
27/38) reveal filamentary configuration with 2 of them being close to 
pure triaxial
$K_{1}\simeq K_{2}$ or ${\cal R}\simeq 1$ (see Table 1: supercluster 31 and 44). The rest of 
the objects (11 or 29$\%$) are estimated to be pancakes, with 3 of them being 
close to pure triaxial (see Table 1: supercluster 17, 18 and 54). 
The mean value of the semi-major supercluster axes is
$\bar{\alpha}_{1}=54.96\pm 22.34h^{-1}$Mpc.
However, for the total supercluster sample we find 
79$\%$ (45/57) filaments and 21$\%$ (12/57) pancakes respectively.
Finally, the histogram in figure 6 presents the total number of 
the rich SDSS superclusters as a function of redshift corresponds 
to a mean density of order 
$\simeq 5.58(\pm 0.74) \times 10^{-6}h^{3}$Mpc$^{-3}$.

\begin{figure}
\mbox{\epsfxsize=8.0cm \epsffile{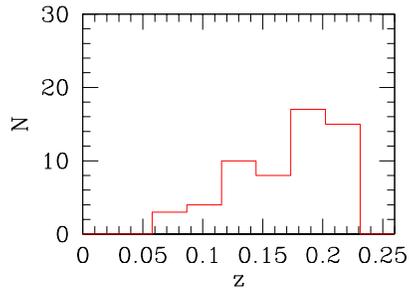}}
\caption{The number of the SDSS superclusters as a function of redshift.} 
\end{figure}
A recent paper by Einasto et al. (2002) examines clusters and superclusters
in the SDSS survey. Superclusters 
are found by these authors utilizing a smoothed apparent magnitude 
limited sample rather than the point distribution of clusters and the 
paper concentrates on measuring the high density regions.
These authors found 43 superclusters,  
35 of which are dubbed as rich ($k\ge 8 $) up 
to redshift 0.2; we find 42 such superclusters.
The smoothing technique owing to the coupling 
between the selection function and the constant radius smoothing
could produce a distorted smoothed density distribution,
especially at large distances (see Gaztanaga \& Yokoyama 1993; 
Basilakos et al. 2001).

\section{Conclusions}
We have studied the morphological parameters of superclusters 
identified in the optical SDSS cluster catalog, up 
to redshift $0.23$, using the
selection features of the SDSS cluster catalog. We have applied 
a constant size percolation radius $R_{\rm pr}$ in order to detect 
superclusters within a distance where the cluster sample is volume
limited. The measure of 
the global supercluster geometry has been based on a 
differential geometry method 
derived by Sahni et al. (1998) and we find that filaments (71$\%$) dominate
over pancakes (29$\%$), in agreement with other recent 
large-scale structure studies.

\section* {Acknowledgements}
This work is jointly funded by the European Union
and the Greek Goverment  in the framework of the programme 'Promotion
of Excellence in Technological Development and Research', project
'X-ray Astrophysics with ESA's mission XMM'.
Finally, I would like to thank the anonymous referee, for his/her
useful suggestions.

\end{document}